\documentclass[letterpaper, 10 pt, conference]{ieeeconf}  

\IEEEoverridecommandlockouts                              
\usepackage{graphics} 
\usepackage{epsfig} 
\usepackage{amsmath} 
\usepackage{esvect}
\usepackage{xcolor}
\usepackage{cite}
\usepackage{amsfonts}

\newtheorem{assumption}{Assumption}
\title{\LARGE \bf
A Minimum-propellant Pontryagin-based Nonlinear MPC for Spacecraft Rendezvous in Lunar Orbit: the Extended Version}

\author{Michele Pagone, Giordana Bucchioni, Francesco Alfino, and Carlo Novara
\thanks{Giordana Bucchioni is with the Department of Information Engineering of University of Pisa, Lungarno Antonio Pacinotti, 43, 56126 Pisa, Italy. Francesco Alfino, Michele Pagone, Carlo Novara are with the Department of Electronics and Telecommunications of Politecnico di Torino,
        Corso Duca degli Abruzzi, 24, 10129 Torino, Italy.
        {\tt\small giordana.bucchioni@ing.unipi.it, francesco.alfino@studenti.polito.it \{michele.pagone, carlo.novara\}@polito.it}}%
}

\newtheorem{remark}{Remark}

\begin{document}

\maketitle
\thispagestyle{empty}
\pagestyle{empty}

\begin{abstract}
We propose a Nonlinear Model Predictive Control approach to spacecraft rendezvous in non-Keplerian Lunar orbits. The approach is based on the Pontryagin Minimum Principle and allows the accomplishment of minimum-propellant maneuvers. The relative motion between the chaser and the target is described by the nonlinear and unstable dynamics of the circular restricted three body-problem. In the proposed formulation, we design a minimum-propellant controller, which leads to a bang-bang behavior of the control signal. Under suitable assumptions, simplified dynamics is employed as prediction model, in order to reduce the complexity of the controller algorithm but, at the same time, without penalizing the controller tracking performance. The proposed approach's effectiveness is validated by a simulation example. 

\end{abstract} 

\section{INTRODUCTION}
In space engineering, the rendezevous and docking (RdV) maneuver is one of the most critical in-orbit operations. The RdV consists in guiding and controlling a spacecraft (SC) (called the chaser) so that it achieves a very
close distance to a passive target while possibly (nearly) nullifying the relative velocity between
the two (see, e.g. \cite{hablani2002, fabrega1997}). The RdV maneuver is essential for space exploration since it is required for complex - manned and unmanned - missions. 

In the last decades, RdV operations have been catching increasing attention within the space research and industry, in the framework of multi-purpose space servicing vehicles for in-orbit servicing and/or active debris removal. Therefore, the new generation of guidance and control systems must be able to guarantee a high technological standard to autonomously perform complex tasks in space such as trajectory planning, obstacle avoidance, and constraints satisfaction, with high accuracy and robustness with respect to external disturbances and model uncertainties. 

Particularly, Moon missions have recently gained interest, as witnessed by the well-known NASA Artemis projects. A key aspect for the success of future missions (e.g. the Gateway mission) is the capability to autonomously perform RdV, which implies the necessity of highly accurate control algorithms to accomplish the last and the most safety-critical phase of the RdV, called close-range rendezvous. As in the classical RdV maneuver, the approach is performed through a series of Hold points \cite{International_RVD} and, in the last kilometers, the Guidance, Navigation, and Control loop is closed with a controller based on the feedback of the relative state measurements. The relative dynamics formulation in the Circular Restricted Three Body Problem (CR3BP) was previously investigated in the literature by \cite{franzini2019relative} and \cite{peng2011optimal}, among others. Many controllers could be used to perform
such maneuver: in \cite{galullo2022closed} and \cite{bucchioni2021guidance} a non-linear State Dependent Riccati Controller is used; a stochastic
robust linear time-variant MPC is employed in \cite{sanchez2020chance} and, in \cite{sato2015spacecraft}, the rendezvous is instead performed
exploiting the natural dynamics.

In this paper, we propose a minimum-propellant NMPC, based on the Pontryagin optimality principle, for a spacecraft rendezvous in Lunar orbit. In detail, we propose a control synthesis for the last phase of the RdV maneuver, being the only closed-loop portion of the full RdV baseline, when the relative chaser-target distance is less than $1~km$ \cite{galullo2022closed}. The chaser relative dynamics (i.e., the plant) and the prediction model used for the controller design and validation are slightly different. Indeed, under suitable assumptions, the relative dynamics formulation of the CR3BP \cite{franzini2019relative} is employed as plant, while a simplified dynamics as prediction model. 
As result, the proposed controller features less algorithmic complexity, without affecting the precision in tracking the reference. Another interesting feature of the presented approach consists of the NMPC algorithm, based on the Pontryagin Minimum Principle (PMP) \cite{pontryagin1962}. An advantage of this approach is the possibility to obtain a control law which is an explicit representation of the state and the costate, even if the system dynamics and/or constraints are nonlinear. On the other hand, one needs to solve a two-points boundary value problem (TPBVP) in order to find the state and the costate. Having an explicit optimal control in terms of state and covector is useful for further and more advanced NMPC schemes, based on the intrinsic characteristics of the dynamics manifolds.

\section{Spacecraft Dynamics in Lunar Orbit as CR3BP}\label{sec:Orbital_Dynamics}
The restricted three-body equations are a commonly used approximation of the dynamics of a spacecraft subject to the Earth and Moon gravity. This hypothesis implies that the spacecraft motion is regulated solely by the influence of two main gravitational masses, called primaries, while the contribution of all the other masses is neglected.
As a result, the dynamics of the spacecraft can be expressed by the following, with $N=2$:
\begin{equation}
    \mathbf{F} = m \; \mathbf{a} = -\sum_{i = 1}^N G \frac{M_i \; m }{||r_i||^2} \bar{\mathbf{r}}_i
    \label{eq: gravity}
\end{equation}
where $M_i$ are the masses of primaries, $r_i$ are the distances of the spacecraft from the bodies, $\bar{\mathbf{r}}_i$ is the versors along the direction of the joining between the gravitational bodies and the moving object, and $G$ is the universal gravity constant.

\begin{figure}
    \centering
    \includegraphics[scale = 0.35]{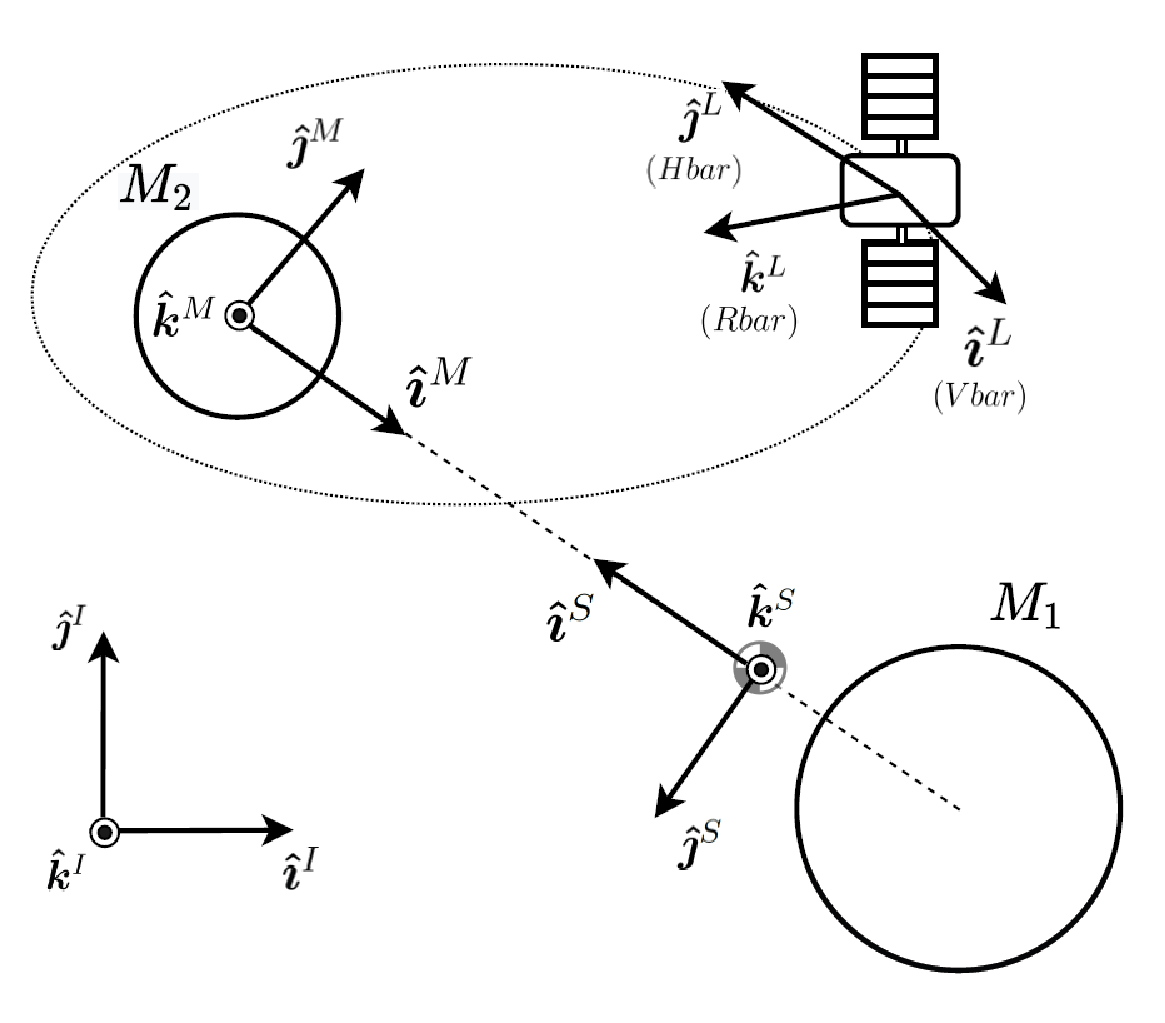}
    \caption{Representation of the reference frame systems employed in the paper.}
    \label{fig:reference_fame}
\end{figure}

The equations of motion can be rewritten under the following assumptions:
\begin{assumption}\label{ass:cr3bp1}
 The system is described with respect to a co-rotating reference frame, called synodic.
$M_1$ and $M_2$ are revolving in a circular motion around the common center of mass.
 All the quantities are normalized according to the characteristics of the system. 
\end{assumption}
Figure \ref{fig:reference_fame} shows the synodic reference frame where the distances of the primaries from the center of mass ($O_s$) are $\mu$ and $1-\mu$, respectively the normalized position of the two primaries (being $\mu$ the system's gravitational parameter, $\mu = M_2/(M_1 + M_2)$) and $\omega$ is the angular velocity of the revolving system, which is normalized to 1.

\begin{figure}
    \centering
    \includegraphics[scale = 0.32]{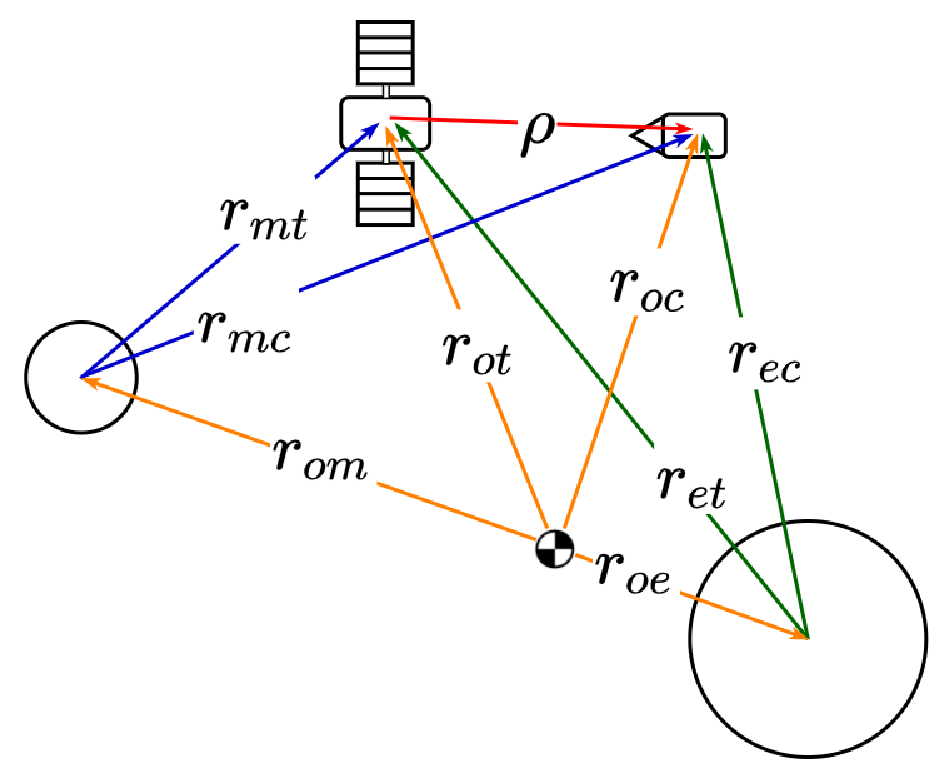}
    \caption{Scheme of vector notations for the chaser relative motion.}
    \label{fig:vector_notation}
\end{figure}

Under these assumptions, the dynamics featuring the CR3BP is described by 
\begin{equation}
    \begin{cases}
      \ddot{x} - 2 \dot{y} = -\dfrac{\partial U}{\partial x} \\
              \ddot{y} + 2\dot{x} = -\dfrac{\partial U}{\partial y} \\
              \ddot{z} = -\dfrac{\partial U} {\partial z} \\
    \end{cases}
    \label{eq:CR3BP}
\end{equation}
where the effective potential, $U$, is given by
\begin{equation}
    U = -{x^2 + y^2 \over 2} - {1 - \mu \over \|\mathbf{r}_{et}\|_2} - {\mu \over \|\mathbf{r}_{mt}\|_2} - {\mu(1 - \mu) \over 2} 
    \label{eq:potential}
\end{equation}
and $[x, y, z]^T$ are the components of the spacecraft position (henceforth, $r_{ot}$) in the synodic reference frame, $r_{mt}$ the vector joining the chaser with the Moon's center, and $r_{et}$ is the chaser-Earth's center position vector (see Figure \ref{fig:vector_notation}).
The CR3BP formulation proves the existence of five equilibrium points, named Lagrangian points, numbered from L1 to L5, around which periodic families of solutions exist. Due to their stability and visibility properties, the most plausible to be used in future missions are the so-called L2 Near Rectilinear Halo orbits \cite{zimovan2017near}.
Therefore, the presented work is focused on this particular family. 
For the sake of clarity, two definitions are given here: the closest point of the
periodic non-keplerian orbit to the Moon is defined as perilune, while the furthest point is named apolune.

\subsection{Rendezvous maneuver and Relative Dynamics}
In the context of the non-keplerian dynamics, the RdV maneuver is not already standardised as in the two-body dynamics, even if some references can be found in \cite{International_RVD}. Mainly, the idea is to replicate the same procedure used around Earth for rendezvous and docking: there are two vehicles, one passive (target), which is located on a non-keplerian orbit, and one active (chaser) which completes the RdV maneuver. 

We remind that the proposed control synthesis copes with the last phase of the RdV operations, since it is the only portion of the maneuver of the full RdV baseline whose controller is in closed-loop. The final RdV phase takes place when the relative chaser-target distance is less than $1~km$. 

The presented work proposes the synthesis of a nonlinear MPC controller to accomplish RdV maneuver in the proximity of the Moon, therefore a local reference system shall be defined in order to model the dynamics of the relative motion of the target with respect to the chaser. Herein, a definition of the Local-Vertical-Local-Horizon (LVLH) is provided, as well as the definition of the relative motion dynamics $L:\{O_t;\{\hat{i}\}^L,\{\hat{j}\}^L,\{\hat{k}\}^L \}$
or equivalently $ L :\{O_t;V_{bar},H_{bar},R_{bar} \}$.
$L$ is centered in the target.
Its basis vectors are defined as:
\begin{equation}\label{lvlh}
 \biggl\{ \mathbf{\hat{j}}^L 
\times    
\mathbf{\hat{k}}^L
\ \ \ 
- \dfrac{{\mathbf{r}}_{mt} \times\dot{\mathbf{r}}_{mt}}{\|{\mathbf{r}}_{mt} \times\dot{\mathbf{r}}_{mt}\|}
\ \ \ 
- \dfrac{{\mathbf{r}}_{mt}}{\|{\mathbf{r}}_{mt}\| }
\biggr\}.
\end{equation}

Therefore, as introduced in \cite{franzini2019relative} the relative position of the chaser $\mathbf{\rho}$ with respect to the target, in the LVLH reference frame, is described by the following nonlinear affine-in-the-input system of differential equations:
\begin{equation}\label{Rho_final}
\begin{split}
\ddot{\mathbf{\rho}} &= -2\bigr[\mathbf{\Omega}_{IL}\bigr]\dot{\mathbf{\rho}} - \bigr[\dot{\mathbf{\Omega}}_{IL}\bigr]\mathbf{\rho}-\bigr[{}\mathbf{\Omega}_{IL}\bigr]^2\mathbf{\rho} + \\ 
& + \mu \biggl( \dfrac{{\mathbf{r}}_{ot}  - {\mathbf{r}}_{om}}{||{\mathbf{r}}_{ot}  - {\mathbf{r}}_{om} ||^3}  - 
\dfrac{{\mathbf{\rho}} + {\mathbf{r}}_{ot} - {\mathbf{r}}_{om}   }{||{\mathbf{\rho}}  + {\mathbf{r}}_{ot} - {\mathbf{r}}_{om}||^3} \biggr)+\\
&+ 
(1 - \mu)   \biggl(   \dfrac{ {\mathbf{r}}_{ot}  - {\mathbf{r}}_{oe}    }{||{\mathbf{r}}_{ot}  - {\mathbf{r}}_{oe} ||^3} 
 -  
\dfrac{ {\mathbf{\rho}}  + {\mathbf{r}}_{ot}  - {\mathbf{r}}_{oe}    }{||  {\mathbf{\rho}} + {\mathbf{r}}_{ot}  - {\mathbf{r}}_{oe} ||^3} \biggr) + u
\end{split}
\end{equation}
where $\mathbf{\Omega}_{IL}$ is the angular velocity of the LVLH frame with respect to an inertial frame, in agreement with the formulation introduced by \cite{franzini2019relative}, and $u$ is the control input. Moreover, the operator $\bigl[\mathbf{\Omega}_{IL}\bigr]$ consists of the skew-symmetric matrix for the vector cross-product. Note that, if not differently specified, all the quantities in \eqref{Rho_final} are expressed in LVLH frame.

\section{NONLINEAR MODEL PREDICTIVE CONTROL SETTING}\label{sec:NMPC}
Henceforth, the relative motion dynamics in \eqref{Rho_final} is simplified under some suitable assumptions, so that it can be employed as prediction model within the NMPC loop. 

\begin{assumption}\label{ass:pred} 
The RdV begins at least 6 hours prior to the passage at the apolune  where the dynamics is slower and the influence of the non-linearities is reduced \cite{bucchioni2021ephemeris}. Moreover, the NMPC prediction interval is a few minutes long.
\end{assumption} 

Given the above assumptions, some terms in \eqref{Rho_final} can be considered constant along the prediction time window. In particular, $\mathbf{\Omega}_{IL}$ is assumed not to vary along the prediction horizon (and so, dropping its time derivative in \eqref{Rho_final}), together with ${\mathbf{r}}_{ot}$, ${\mathbf{r}}_{om}$, and ${\mathbf{r}}_{oe}$. This simplifies the overall complexity of the optimal control problem, without reducing the generality and the optimality of the NMPC solution. In few words, we have considered that, along the prediction time window, the spacecraft, the chaser, and the Moon displacements with respect to Earth are negligible. To sum up, the dynamics in \eqref{Rho_final} is employed as the plant, fed with the optimal control law. The simplified dynamics coming from Assumption \ref{ass:pred} is used as the prediction model of the system. 

By accounting for the relative motion in \eqref{Rho_final}, together with Assumption \ref{ass:pred}, the chaser-target dynamics is described by a set of nonlinear time-invariant affine-in-the-input differential equation as:
\begin{equation}\label{eq:dyn}
\dot{x}(t) = f(x(t)) + g(x(t))u(t), ~~ t \geq 0,
\end{equation}
where $x(t) \in \mathbb{R}^{n_x}$, $u(t) \in \mathbb{R}^{n_u}$ are, respectively, the state (chaser relative position and velocity), and the input (thrust acceleration) vectors at time $t$. Moreover, $f \in \mathcal{C}^1(\mathbb{R}^{n_x} \rightarrow \mathbb{R}^{n_x}$ and $g \in \mathcal{C}^1(\mathbb{R}^{n_x} \rightarrow \mathbb{R}^{n_x \times n_u})$.

At each $k$-th time instant, the state is measured in real-time - with a sampling time $T_S$ - and a prediction $\hat{x}$ of the system state is performed over a finite time interval $[t_k, t_k + T_p]$, where $T_p$ is the prediction horizon. The prediction is obtained by integration of \eqref{eq:dyn}. Based on the state measurement, we compute the control input $u(t)$ along the prediction window. To this end, fixed a prediction horizon $T_p \geq T_s$ and solve the Bolza-type optimization problem:
\begin{equation} \label{eq:functional}
\begin{split}
& u^* = \arg \min J\bigl(x(t),u(t)\bigr) \\
& \text{subject to:} \\
& \dot{\hat{x}}(\tau) = f(\hat{x}(\tau)) + g(\hat{x}(\tau)) \hat u(\tau),~~\hat{x}(t_k) = x(t_k) \\
& \hat{x}(\tau) \in \mathcal{X}\subset\mathbb{R}^{n_x}, ~\hat u(\tau) \in \mathcal{U}\subset\mathbb{R}^{n_u}, ~~\forall \tau \in [t_k,t_k+T_p] ,\\
& \hat u(\cdot)\in \mathcal{PC}\bigl([t_k, t_k + T_p]\bigr).
\end{split}
\end{equation}
$\mathcal{X}$ and $\mathcal{U}$ are set describing possible constraints on the state and input, respectively, while $\mathcal{PC}\bigl([t_k, t_k + T_p]\bigr)$ is the space of piece-wise continuous functions. Upon finding the optimal solution $(\hat x^*,\hat u^*)$, the control input $u(t) \in [t_k,t_{k+1}]$ is defined as the constant value
\begin{equation}\label{eq:control-discrete}
u(t)\equiv\hat u^*(t_k)\quad\forall t\in [t_k,t_{k+1}).
\end{equation}
This operation is performed at each sampling instant $t_k$, where $k=0,1,\ldots$.
Note that, as in the usual scenario for the MPC setting, the receding horizon strategy is employed.

In the framework of the space RdV, a particular kind of cost function - which promotes the minimum propellant expenditure - is employed. As already presented in \cite{pagone2021}, in the following, we minimize the $\mathcal{L}_1$ norm of the input, since it effectively leads to a minimum-propellant controller \cite{ross2004}. As a consequence, the resulting input signal will be bang-bang in time. Associating to each solution $\hat{x}$ the corresponding tracking error $\tilde{x}(\tau) \doteq \hat{x} - x_r$ (where $x_r$ is the reference signal), the following functional is introduced:
\begin{equation}\label{eq:J}
\begin{split}
    J &= \int_{t_k}^{t_k+T_p}  \bigl(\tilde{x}^T(\tau)\mathbf{Q}\tilde{x}(\tau) +
\|\mathbf{R}\hat u(\tau)\|_2\bigr)~\mathrm{d}\tau~ ~ \\
&+ \tilde{x}^T(t_k + T_p)\mathbf{P}\tilde{x}(t_k + T_p) 
\end{split}
\end{equation}
where $\mathbf{Q},\mathbf{P} \geq 0 \in \mathbb{R}^{n_x \times n_x}$ and $\mathbf{R} > 0 \in \mathbb{R}^{n_u \times n_u} $ are suitable diagonal matrices.

For the application at hand, the admissible control set is defined as $\mathcal{U} = \{u \in \mathbb{R}^{n_u}: \|u\|_2 \leq u_{max} \}$ where $u_{max}$ is the maximum thrust acceleration deliverable by the engines. Unlike the input, in this preliminary work, the state is considered unconstrained, then, we assume that $\mathcal{X} \equiv \mathbb{R}^{n_x}$. Note that, the integration of state constraints can be carried out by following the methodology described in \cite{pagone2021} and \cite{pagone2022}.


\begin{remark}
Stability of nonlinear MPC algorithms is a classical problem, which is still far from being completely solved. The easiest way to ensure stability consist of adjoining the optimal control problem with a terminal constraint $x(t_k + T_p) \in \Omega$ where $\Omega \subset \mathcal{X}$ is some properly constrained set (see, e.g., \cite{chen1998, mayne1990}). On the other hand, stability can also be ensured (at least locally) by eliminating the terminal constraints. This latter choice gives a number of advantages. Indeed, in presence of such a constraint, the optimal control problem is feasible only if the equilibrium (or its predefined neighborhood) is reachable from $x(0)$ in time $T_p$, which leads to another sophisticated problem of tuning the prediction horizon. Some results of NMPC stability without terminal constraints can be found in \cite{alamir2018} and \cite{grimm2005}. In practice, one can choose $\mathbf{P}$ in such a way that all solutions starting from the set of all initial conditions, from which the reference is reachable, are uniformly bounded and, furthermore, converge to a predefined neighborhood of the equilibrium. In this sense, we can make the NMPC stable by choosing the terminal weight large enough: the more is larger the terminal weight, the tighter is the invariant set - centered at the reference - bounding the solution.
\end{remark}

\subsection{Pontryagin-based NMPC Solution}
To solve the optimal control problem \eqref{eq:functional}, we employ the well-established Pontryagin principle \cite{pontryagin1962}. The necessary condition for optimality requires the introduction of the Hamiltonian
\begin{equation}\label{eq:ham}
    H(x,u,\lambda) = \tilde{x}^T\mathbf{Q}\tilde{x} + \|\mathbf{R}u\|_2 + \lambda^T[f(x) + g(x)u] \in \mathbb{R}.
\end{equation}
where $\lambda \in \mathbb{R}^{n_x}$ is the vector of costates (or covector). 

Denoting, for brevity, $t^f\doteq t_k+T_p$, the necessary conditions of optimality in \eqref{eq:functional} are as follows~\cite{bryson1975}. If $(\hat x^*,\hat u^*)$ is an optimal solution, then a costate function $\lambda^*:[t_k,t^f]\to\mathbb{R}^{n_x}$ exists such that
\begin{gather}
\dot\lambda^*(t)=-\nabla_xH(\hat x^*(t),\hat u^*(t),\lambda^*(t)), \label{eq:el_eq}\\
\lambda^*(t^f)=\nabla_x \tilde{x}^T(t^f)\mathbf{P}\tilde{x}(t^f),\label{eq:bc_new}\\
H(\hat x^*(t),\hat u^*(t),\lambda^*(t))=min_{u\in U_C}H(\hat x^*(t),u,\lambda^*(t))\label{eq:opt-u}\\
\forall t\in[t_k,t^f],\quad t^f\doteq t_k+T_p.
\end{gather}
Adding constraints~\eqref{eq:functional} to this system, one obtains a two-point boundary value problem (TPBVP) with $2n$ scalar differential equations for
the vector-function $(\hat x^*(t),\lambda^*(t))\in\mathbb{R}^{2n}$, $n$ scalar boundary conditions at time $t_k$ (namely, $\hat x^*(t_k)=x_k$) and $n$ boundary conditions at time $t^f$.

\begin{remark}
Considering the boundary conditions on the state, at time $t = t_k$, the state values cannot be chosen arbitrary: the continuity between two successive sampling steps must be ensured, whereas, at $t_k + T_p$, no further conditions are required. Then, if the variable is assigned at $t_k$, the corresponding covector value is free (see, e.g., \cite{bryson1975}). As for the boundary conditions on time (i.e., transversality conditions), both the initial and final time are assigned (trivially, the initial and final time must coincide with the borders of the prediction horizon), then, the
corresponding values of the Hamiltonian are free \cite{bryson1975}.
\end{remark}

\subsection{Rendezvous Optimal Control Problem}
In order to present the Pontryagin-based solution for the RdV dynamics, let $\lambda_r$ and $\lambda_v$ be defined as the co-vectors associated with the chaser relative position $\rho$ and the relative velocity $\dot{\rho}$, respectively. 
Hence, by considering \eqref{eq:ham}, together with cost function \eqref{eq:J}, the Hamiltonian for the RdV optimal control problem is
\begin{equation}
H_{RdV} = \tilde{x}^T\mathbf{Q}\tilde{x} +
\|\mathbf{R}u\|_2 + \lambda_r^T \dot{\rho} + \lambda_v^T\ddot{\rho}.
\end{equation}
Whereby, by taking into account \eqref{Rho_final}
\begin{equation}
\begin{split}
H_{RdV} &= \tilde{x}^T\mathbf{Q}\tilde{x}  + \|\mathbf{R}u\|_2 + \lambda_r^T \dot{\rho}  +\lambda^T_v \biggl[-2\bigr[\mathbf{\Omega}_{IL}\bigr]\dot{\mathbf{\rho}} +\\
&- \bigr[\dot{\mathbf{\Omega}}_{IL}\bigr]\mathbf{\rho}  
-\bigr[{}\mathbf{\Omega}_{IL}\bigr]^2\mathbf{\rho} +\mu\ell(\rho) + (1-\mu)\kappa(\rho) + u\biggr]
\end{split}
\end{equation}
where $\ell(\rho)$ and $\kappa(\rho)$ are defined as
\begin{equation}
    \begin{split}
        \ell(\rho) &= \biggl( \dfrac{{\mathbf{r}}_{ot}  - {\mathbf{r}}_{om}}{||{\mathbf{r}}_{ot}  - {\mathbf{r}}_{om} ||^3}  - 
\dfrac{{\mathbf{\rho}} + {\mathbf{r}}_{ot} - {\mathbf{r}}_{om}   }{||{\mathbf{\rho}}  + {\mathbf{r}}_{ot} - {\mathbf{r}}_{om}||^3} \biggr),\\
\kappa(\rho) &= \biggl(   \dfrac{ {\mathbf{r}}_{ot}  - {\mathbf{r}}_{oe}    }{||{\mathbf{r}}_{ot}  - {\mathbf{r}}_{oe} ||^3} 
 -  
\dfrac{ {\mathbf{\rho}}  + {\mathbf{r}}_{ot}  - {\mathbf{r}}_{oe}    }{||  {\mathbf{\rho}} + {\mathbf{r}}_{ot}  - {\mathbf{r}}_{oe} ||^3} \biggr).
    \end{split}
\end{equation}
In order to explicit the optimal control law, let the Hamiltonian be slightly modified. Define with $\Gamma \doteq \|u\|_2$ the magnitude of the thrust acceleration and $\bar{u}$ the thrust unit vector. Consider, then, the cost rate $\|\mathbf{R}u\|_2$, it is well known that $\|\mathbf{R}u\|_2 \leq \|\mathbf{R}\|_2\|u\|_2$. Nevertheless, if considering that $\mathbf{R}$ is a diagonal definite positive matrix, whose entries are all equal (quite a common setting in the aerospace field), the $\|\mathbf{R}u\|_2 = \|\mathbf{R}\|_2\|u\|_2$ holds. Then, the Hamiltonian turns into
\begin{equation}
\begin{split}
 H_{RdV} &= \tilde{x}^T\mathbf{Q}\tilde{x} + \|\mathbf{R}\|_2\Gamma + \lambda_r^T \dot{\rho} +\lambda^T_v \biggl[-2\bigr[\mathbf{\Omega}_{IL}\bigr]\dot{\mathbf{\rho}} +\\
 &- \bigr[\dot{\mathbf{\Omega}}_{IL}\bigr]\mathbf{\rho} -\bigr[{}\mathbf{\Omega}_{IL}\bigr]^2\mathbf{\rho} +\mu\ell(\rho) + (1-\mu)\kappa(\rho) + \Gamma \bar{u}\biggr].
\end{split}
\end{equation}

We recall now the notion of primer vector, denoted by $p$, introduced by \cite{lawden1963}. The velocity covector represents the engines optimal fire direction so that $p \doteq -\lambda_v$. Hence, $\bar{u} = p/P$, being $P = \|p\|_2 = -\lambda_v^T \bar{u}$ the primer vector magnitude. By the previous notion, the Hamiltonian turns into
\begin{equation}\label{eq:ham_final}
\begin{split}
H_{RdV} &= \tilde{x}^T\mathbf{Q}\tilde{x} +
\|\mathbf{R}\|_2\Gamma + \lambda_r^T \dot{\rho} +\lambda^T_v \biggl[-2\bigr[\mathbf{\Omega}_{IL}\bigr]\dot{\mathbf{\rho}} +\\
 &- \bigr[\dot{\mathbf{\Omega}}_{IL}\bigr]\mathbf{\rho} -\bigr[{}\mathbf{\Omega}_{IL}\bigr]^2\mathbf{\rho} +\mu\ell(\rho)\\
 &+ (1-\mu)\kappa(\rho)\biggr] - P\Gamma = \\
 &= -(P - \|\mathbf{R}\|_2)\Gamma  + \lambda_r^T \dot{\rho} +\lambda^T_v \biggl[-2\bigr[\mathbf{\Omega}_{IL}\bigr]\dot{\mathbf{\rho}} +\\
 &- \bigr[\dot{\mathbf{\Omega}}_{IL}\bigr]\mathbf{\rho} -\bigr[{}\mathbf{\Omega}_{IL}\bigr]^2\mathbf{\rho} +\mu\ell(\rho) + (1-\mu)\kappa(\rho)\biggr].
\end{split}
\end{equation}
Hence, the Hamiltonian must be minimized over the choice of the thrust magnitude $\Gamma$, which appears linearly in \eqref{eq:ham_final}. Thus, the optimal control problem solution would lead to an input signal with an infinite magnitude. Nevertheless, if the admissible input set is bounded, the minimization of the Hamiltonian will depend only on the algebraic sign of the $\Gamma$ coefficient, which, in aerospace literature is defined as switching function $\Upsilon = P - \|\mathbf{R}\|_2$. The sign of $\Upsilon$ defines the policy for the engines power on/off and the thrust is allowed to assume only the maximum or zero value. The direction of the thrust is driven by $p = -\lambda_v$. Therefore, the optimal control policy is:
\begin{equation}\label{eq:opt_con_bang}
u^* = 
\begin{cases}
\Gamma_{max}\dfrac{p}{\|\lambda_v\|_2} ~~ &\mathrm{if} ~~ \Upsilon > 0, \\
0 ~~&\mathrm{if} ~~ \Upsilon \leq 0.
\end{cases}
\end{equation}
\begin{remark}
Note that, when $\Upsilon$ vanishes the problem of singular control must be tackled. In these situations, the optimal control must be searched to find an explicit expression of u by nullifying the time derivatives of $\nabla_uH$, until $u$ appears (see, e.g. \cite{zelikin2005}). As consequence, the optimal control does not lie anymore on the boundary of $\mathcal{U}$ but it can assume any value inside the set. Nevertheless, for the applications dealt with in this work, it is important to stress that the bang-bang control problem is also driven by the necessity to cope with some technological limitations of the actuators whose output must have a switch on/off behaviour. For this reason, in this peculiar application, a suitable - but sub-optimal - choice to deal with the singular control is to set the corresponding value to zero.
\end{remark}

We conclude the PMP-based NMPC solution by presenting the so-called Euler-Lagrange equations which describe the time variation of the costate as in \eqref{eq:el_eq}:
\begin{equation}
\dot{\lambda} = 
\begin{bmatrix}
{\dot{\lambda}}_r \\ \dot{\lambda}_v 
\end{bmatrix} -
2 \mathbf{Q}  \tilde{x}
\end{equation}
where
\begin{subequations}
\begin{equation}\label{eq:lr}
\begin{split}
\dot{\lambda}_r &= \biggl[\bigr[\dot{\mathbf{\Omega}}_{IL}\bigr]^T\bigr[\dot{\mathbf{\Omega}}_{IL}\bigr]^T + \sum_{i=1}^2\dfrac{\textrm{M}_i}{\|{\rho}  + \mathbf{r}_{ot} - \mathbf{r}_{oi}\|^3} \cdot\\
&\cdot\biggl(\mathbf{I} -3 \dfrac{(\rho +\mathbf{r}_{ot} - \mathbf{r}_{oi})^T(\rho +\mathbf{r}_{ot} - \mathbf{r}_{oi})}{\|\rho +\mathbf{r}_{ot} - \mathbf{r}_{oi}\|^2}
\biggr)\biggr] \lambda_v
\end{split}
\end{equation}
\begin{equation}\label{eq:lv}
\dot{\lambda}_v  = -\lambda_r + 2\bigr[\mathbf{\Omega}_{IL}\bigr]^T  \lambda_v.
\end{equation}
\end{subequations}
where $\mathbf{I}$ is the identity matrix, $\textrm{M} = \{\mu, 1-\mu\}$, and $r_{oi} = \{\mathbf{r}_{om}, \mathbf{r}_{oe} \}$. Note that, referring to \eqref{eq:lr} and \eqref{eq:lv}, we remind that the time variation of the costate is evaluated in the time window $[t_k, t_k + T_p]$. Then, according to Assumption \ref{ass:pred}, $\mathbf{\Omega}_{IL}$, $\mathbf{r}_{ot}$, $\mathbf{r}_{oe}$, and $\mathbf{r}_{om}$ are constant.

\section{SIMULATION ANALYSIS}\label{sec:simulations}
\begin{figure}[thpb]
 \centering
\includegraphics[scale=0.5]{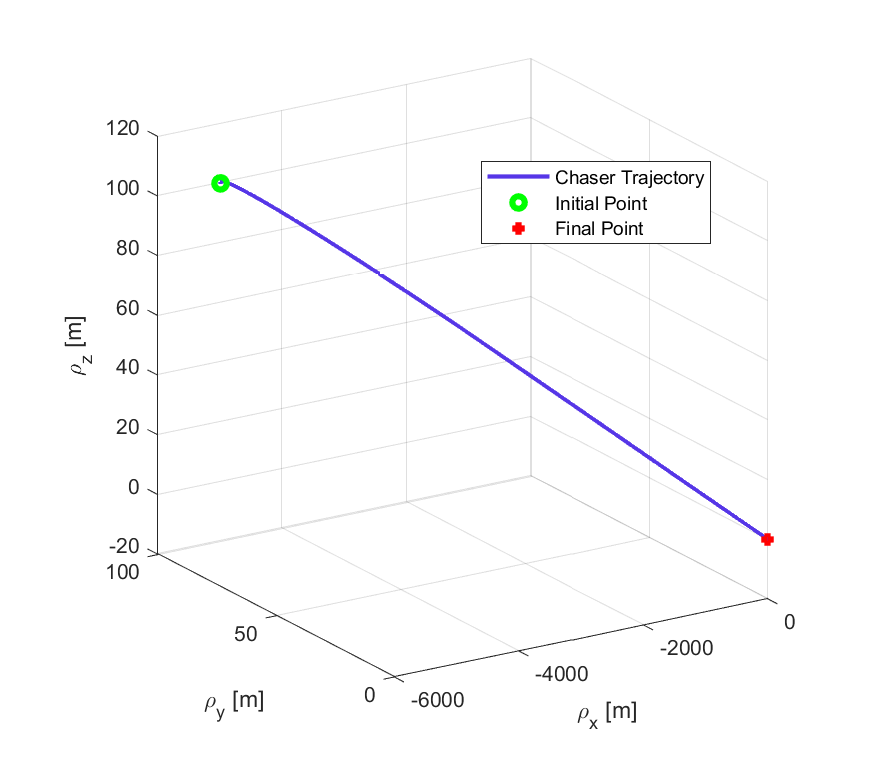}   
\caption{Trajectory of the chaser SC from the initial point (green circle) to the target (red dot) in the chaser/target relative reference frame. }
\label{fig:traj}
  \end{figure}

The proposed control technique is then applied in a specific test case: for realistic purposes, the target orbit was selected to be the same as the Lunar Gateway (mission Artemis IV) \cite{zimovan2017near}, using as numerical values for the simulation of the environment: $\mu = 0.01215$, Earth-Moon distance $L = 384400.0 km$ and the Earth-Moon synodic revolution period $ T = 2360591.424 s$. The rendezvous maneuver is accomplished at least 6 hours before reaching the aposelene to mitigate the effect of the non-linear dynamics, in particular, the focus is posed on the close RdV, therefore the chaser shall be located on a hold point (quasi-zero relative velocity) at $-5~km$ along V-bar. 
The time to accomplish the full mission shall be less than 4h.
For the application at hand, the initial conditions are set as $\rho_0 = [-5, 0.1, 0.1]^T~[km]$ and $\dot{\rho}_0 = [2e-5, 2e-5, 2e-5]^T~[km/s]$ while the reference is a constant zero vector both for position and velocity, except for the components $\rho_x$ set at $-5~m$, in agreement with \cite{bucchioni2021guidance}. Concerning the NMPC parameters, we have that $T_s = 2~s$ and $T_p = 45\cdot T_s$, while, the cost function matrices are $\mathbf{R} = \mathbf{I}_{3 \times 3}$ (whereas $\mathbf{I}$ is the identity matrix), $\mathbf{P} = \mathrm{diag}(8.05e10, 8.05e10, 8.05e10, 1, 1, 1)$, and $\mathbf{Q} = \mathrm{diag}(5e14, 5e14, 5e14, 9e7, 9e7, 9e7)$, chosen by a trial and error procedure. Finally, the thrust acceleration provided by the engines is allowed to vary within the set $\mathcal{U} = \{ u: \|u\|_2 \leq 0.02~m/s^2\}$ which approximately corresponds to a maximum thrust force of $10~N$.
We are now in position to present the outcomes of the simulations. The orbital simulator and the NMPC algorithm are implemented in the Matlab/Simulink environment. The optimal control problem, cast as a TPBVP, is solved by means of \textit{bvp5c} Matlab function. In Figure \ref{fig:traj}, the full RdV trajectory is represented in the chaser/target relative reference frame, showing the motion of the chaser from the initial point towards the reference point.
  
\begin{figure}[thpb]
 \centering
\includegraphics[scale=0.52]{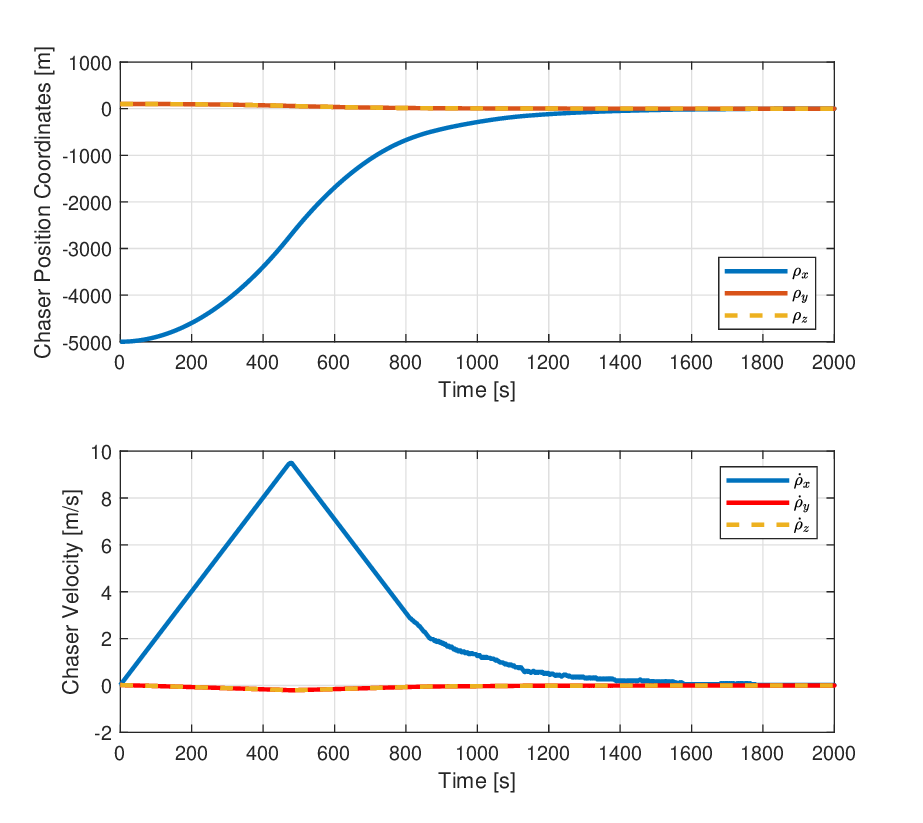}   
\caption{Time evolution of the chaser position/velocity components. The components along the z-axis are dashed in order to visualize the overlapping with the y-axis ones.}
\label{fig:pos_vel}
  \end{figure}

Concerning the tracking performance of the controller, Figure \ref{fig:pos_vel} shows how both the chaser position and velocity components own excellent convergence properties. Indeed, whereas the final tracking error on x-axis is below $1~m$, the displacements on y/z-axis are in the order of a few millimetres. Finally, the residual velocity components are all below the threshold of $cm/s$. Note that, small oscillations of the chaser nearby the reference can make the two SC collide. In order to avoid collision a keep-out-zone can be included within the NMPC optimization problem, with the same methodology proposed in \cite{pagone2021}. The implementation of such constraint is a topic of on-going work.

\begin{figure}[thpb]
 \centering
\includegraphics[scale=0.52]{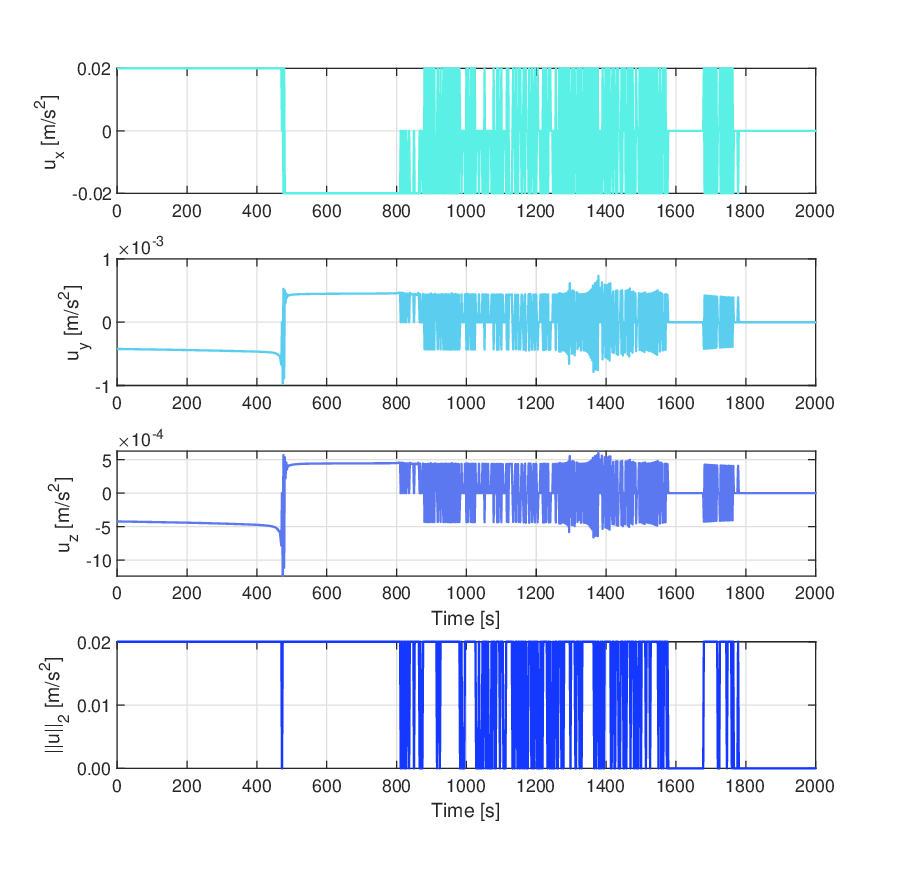}   
\caption{Thrust delivered by the chaser engines along the trajectory in terms of components and total magnitude. The bang-bang behaviour is evident in the last subplot where $\|u\|_2$ is plotted. The engines can assume only an on-off policy.}
\label{fig:thrust}
  \end{figure}

By evaluating the input time history, it is useful to stress the following point. According to \eqref{eq:opt_con_bang}, the bang-bang behaviour of the thrust activity is evident in the last subplot of Figure \ref{fig:thrust}. Indeed, we remind bang behaviour is meant to refer to the magnitude acceleration and not component-wise. For this reason, the single components of the thrust can assume any value within the input set, being the input constraint always satisfied. Moreover, even though the thrust acceleration evolution presents a high-frequency behaviour, the issue can be mitigated through a proper control dispatch during the S/C engines configuration design. To conclude, the overall impulse $I_u = \int_{t_0}^{t_F}\|u\|_2 \mathrm{d}t$ delivered by the engines is $I_u = 28~m/s^2 \cdot s$. This latter result is comparable with the results obtained with SDRE - State Dependent Riccati Equations - controller formulated in \cite{bucchioni2021guidance}.

\begin{figure}[thpb]
 \centering
\includegraphics[scale=0.56]{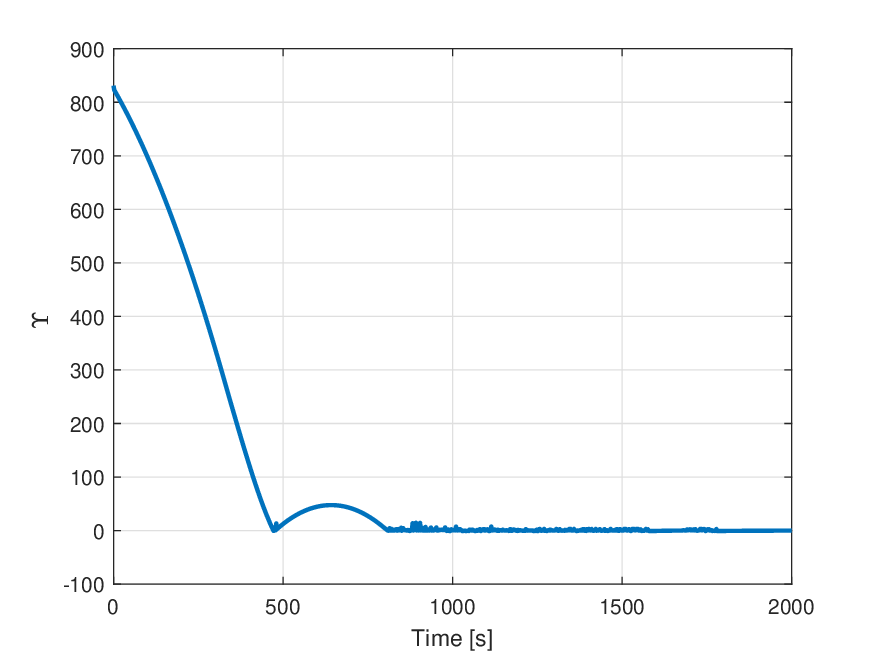}   
\caption{Switching function $\Upsilon$ along the chaser approach. }
\label{fig:SF_full}
  \end{figure}

\begin{figure}[thpb]
 \centering
\includegraphics[scale=0.56]{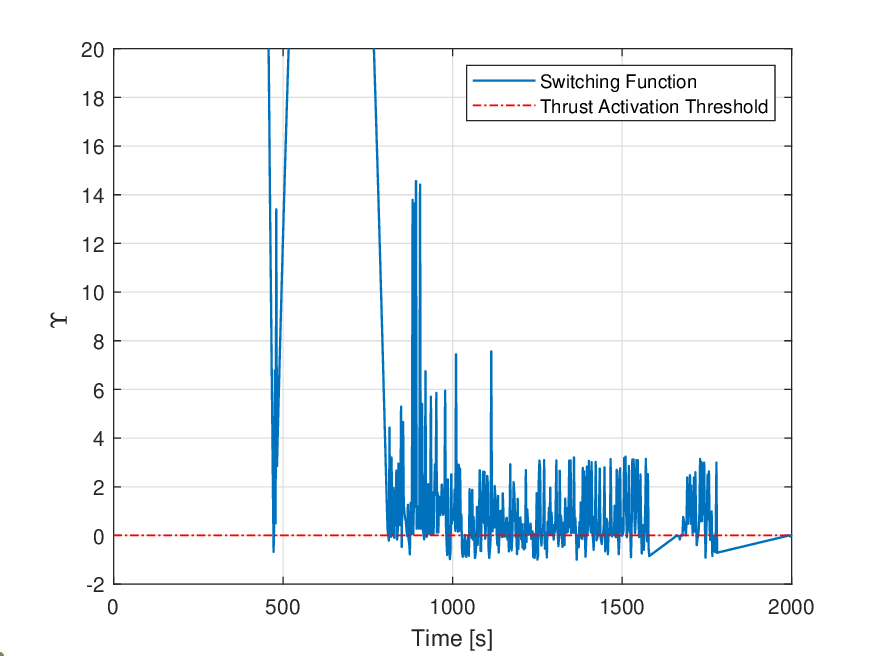}   
\caption{Zoom of the switching function $\Upsilon$ along the chaser approach. The switching activation policy in plotted in blue, while the zero-line, i.e., the engines activation threshold in red.}
\label{fig:SF_zoom}
  \end{figure}

It is also worth to highlight the behavior of the switching function $\Upsilon$ along the full maneuver of the chaser in Figure \ref{fig:SF_full}. Nevertheless, we refer to Figure \ref{fig:SF_zoom} for a more detailed analysis, were we show a zoom of $\Upsilon$ in the region where it crosses the zero. By comparing the trend of $\Upsilon$ in Figure \ref{fig:SF_zoom} with the one of the thrust magnitude in Figure \ref{fig:thrust}, it is evident the correspondence of the segment in which $\Upsilon \leq 0$ with the ones where the engines are switched off.

\end{document}